\documentclass[graybox]{svmult}

\usepackage{bm}
\usepackage{type1cm}        
\usepackage{makeidx}         
\usepackage{graphicx}        
\usepackage{multicol}        
\usepackage[bottom]{footmisc}
\usepackage{float}

\usepackage{newtxtext}       %
\usepackage{newtxmath}       

\DeclareMathOperator{\Tr}{Tr}

\usepackage{booktabs}
\usepackage{url}
\usepackage{multirow,bigdelim}

\makeindex             

\begin{document}

\title*{Flow-based Community Detection in Hypergraphs}
\author{Anton Eriksson, Timoteo Carletti, Renaud Lambiotte, Alexis Rojas, and Martin Rosvall}
\institute{Anton Eriksson \at Ume{\aa} University, Sweden, \email{anton.eriksson@umu.se}
\and Timoteo Carletti \at University of Namur, Belgium, \email{timoteo.carletti@unamur.be}
\and Renaud Lambiotte \at University of Oxford, United Kingdom, \email{renaud.lambiotte@maths.ox.ac.uk}
\and Alexis Rojas \at Ume{\aa} University, Sweden, \email{alexis.rojas-briceno@umu.se}
\and Martin Rosvall \at Ume{\aa} University, Sweden, \email{martin.rosvall@umu.se}
}
%
%
\maketitle

\abstract*{
}

\abstract{
To connect structure, dynamics and function in systems with multibody interactions, network scientists model random walks on hypergraphs and identify communities that confine the walks for a long time.
The two flow-based community-detection methods Markov stability and the map equation identify such communities based on different principles and search algorithms. But how similar are the resulting communities? We explain both methods' machinery applied to hypergraphs and compare them on synthetic and real-world hypergraphs using various hyperedge-size biased random walks and time scales. We find that the map equation is more sensitive to time-scale changes and that Markov stability is more sensitive to hyperedge-size biases.
\newline\newline
{\small With new introduction, experiments, and conclusion sections, this text reuses some text from Carletti, T., Fanelli, D., Lambiotte, R.: Random walks and community detection in hypergraphs. J.\ Phys.\ Complex.\ 2 015011 (2021) and Eriksson, A., Edler, D., Rojas, A., Rosvall, M.: Mapping flows on hypergraphs: How choosing random-walk model and network representation matters for community detection. Comm.\ Phys.\ (2021).}
}

\section{Introduction}
\label{sec:introduction}

Researchers model and map flows on networks to identify important nodes and detect significant communities \cite{brin1998anatomy,simonsen2004diffusion,guimera2005functional,rosvall2008maps,delvenne2010stability,mangioni2018multilayer}. From small to large system scales, random walk-based methods help to uncover the inner workings of networks ~\cite{boccaletti2006complex,fortunato2010community}. When standard network models with dyadic relations between pairs of nodes fail to adequately represent a system's interactions, researchers turn to higher-order models of complex systems~\cite{lambiotte2019networks,battiston2020networks}, including multilayer networks~\cite{mucha2010community,de2013mathematical,kivela2014multilayer,de2016physics} for multitype interactions, memory networks~\cite{rosvall2014memory,scholtes2014causality,xu2016representing} for multistep interactions, and simplicial complexes~\cite{parzanchevski2017simplicial,salnikov2018simplicial,iacopini2019simplicial,schaub2020random} and hypergraphs~\cite{zhou2007learning,chitra2019random,carletti2020random,carletti2020randomwalks} for multibody interactions.

While several methods can identify flow-based communities in multilayer~\cite{mucha2010community,de2015identifying,jeub2017local} and memory~\cite{rosvall2014memory,scholtes2014causality,xu2016representing} networks with higher-order Markov dynamics, researchers have focused on combinatorial methods to identify communities in  hypergraphs~\cite{angelini2015spectral,chien2018community,NIPS2017_a50abba8,kaminski2019clustering,ke2019community,chodrow2021hypergraph} and only recently begun to unravel flow-based community structures associated with random walks guided by hyperedges sizes~\cite{carletti2020randomwalks}.
Two such methods are Markov stability and the map equation. 
Both algorithms exploit random walkers' tendency to stay unexpectedly long times in assortative communities, albeit in different ways. 
Markov stability measures the fraction of random walks that reside in the community where they started after time $t$ compared with stationarity.
Instead, the map equation measures the shortest possible modular codelength required to describe the random walker on the network with given communities. 
Also their optimisation algorithms differ and it remains unclear how their detected communities in hypergraphs compare.
 
For Markov stability, we have previously analysed random walks on hypergraphs with a parameter controlling the bias of the dynamics towards small or large hyperedges and identified widely differing communities \cite{carletti2020randomwalks}. For the map equation and its optimisation algorithm Infomap \cite{infomap}, we have derived and clustered unipartite, bipartite, and multilayer network representations of hypergraph flows with different advantages \cite{eriksson2021mapping}. Both papers highlight that the network and the research question should decide which model to use, but did not compare the two methods. For example, when modelling the flow of ideas with random walks in a co-authorship hypergraph, how does the organisation of authors in communities, at scales from research groups to research areas, change with community-detection method?

We compare Markov stability and the map equation applied to random walks on hypergraphs with hyperedge-size bias. After explaining the random-walk model and the flow-based community-detection methods using a schematic hypergraph for illustration, we consider experiments on three real-world hypergraphs: a zoo hypergraph with $101$ nodes, a collaboration hypergraph with $361$ nodes, and a fossil-record hypergraph with $13,276$ nodes. We find that the map equation responds faster to time-scale changes and that Markov stability is more sensitive to hyperedge-size biases.

\begin{figure*}[!thb]
 \centering
 \includegraphics[width=1\textwidth]{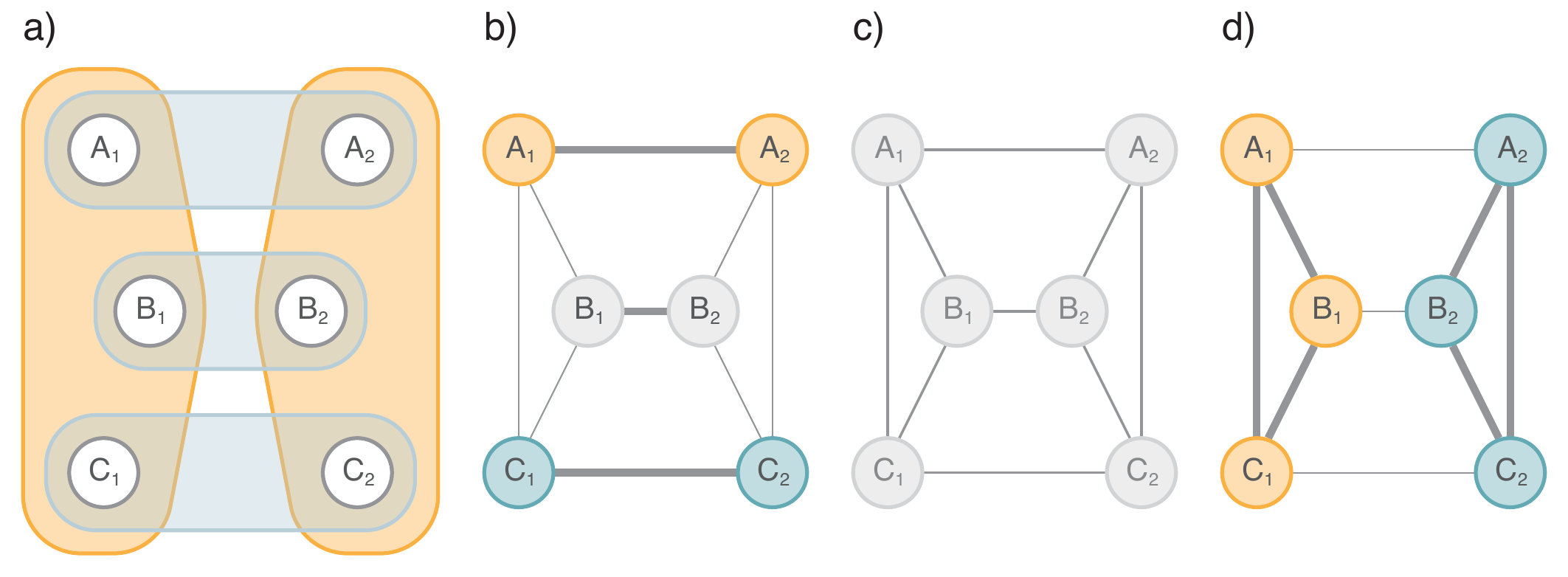}
   \caption{A schematic hypergraph in (a) projected into three different weighted networks with hyperedge-size bias $\sigma=-1$ in (b), $0$ in (c), and $1$ in (d).\label{fig:hypergraphRepresentations}}
\end{figure*}

\section{Random walks on hypergraphs}
\label{sec:rwh}

We consider hypergraphs $\mathcal H(V,E)$ with $n$ nodes $V=\{v_1,\dots,v_n\}$ and $m$ hyperedges $E=\{E_1,\dots,E_m\}$. Each hyperedge $E_\alpha\subset V$ is an unordered collection of nodes. A hypergraph can be encoded by its incidence matrix $e_{i \alpha}$, where we use Roman indexes for nodes and Greek indexes for hyperedges:
\begin{equation}
\label{eq:incid}
e_{i \alpha}=\begin{cases} 1 &\text{$v_i\in E_{\alpha}$}\\
0 & \text{otherwise}\, .
\end{cases}
\end{equation}

The $n\times n$ adjacency matrix of the hypergraph is $\mathbf{A}=\mathbf{e}\mathbf{e}^{\top}$, whose entry $A_{ij}$ represents the number of hyperedges containing both nodes $i$ and $j$ with the diagonal elements set to zero. We also define the $m\times m$ hyperedge matrix $\mathbf{B}=\mathbf{e}^{\top}\mathbf{e}$, whose entry $B_{\alpha \beta}$ counts the number of nodes the two hyperedges share in the original hypergraph, $E_{\alpha}\cap E_{\beta}$. $\mathbf{B}$ corresponds to the weighted adjacency matrix of the dual hypergraph, where hyperedges of the original hypergraph become nodes of the new structure. 

A random walk process on a hypergraph can be defined by the transition probability, $T_{ij}$, allowing the walker to move across any pair of nodes $(i,j)$. The resulting continuous-time Markov process can be defined by
\begin{equation}
\dot{p}_i(t)=\sum_j p_j(t)T_{ji} - \sum_j p_i{(t)}T_{ij}\, ,
\end{equation}
where ${p}_i(t)$ is the probability of finding the walker on node $i$ at time $t$. As often assumed when dealing with Markov processes, $\mathbf{p}=(p_1,\dots,p_n)$ is a row vector.
With normalised transition probabilities such that $\sum_j T_{ij}=1$, we rewrite the continuous-time Markov process as
\begin{equation}
\dot{p}_i=\sum_j p_j(T_{ji} - \delta_{ij})=-\sum_j p_jL_{ji}\, ,
\label{eq:ctrw} 
\end{equation}
where $L_{ij} := \delta_{ij}-T_{ij}$ is the random-walk Laplace operator.

Assigning weights to nodes and hyperedges biases the transition probabilities and leads to different probability flows between nodes~\cite{zhou2007learning,carletti2020random,carletti2020randomwalks}.
\begin{svgraybox}
\textbf{Fixed weights and independent moves.}~\cite{zhou2007learning}\\
Setting a weight $\omega(E_\alpha)$ to each hyperedge $E_\alpha$ and a weight $\gamma_{E_\alpha}(i)$ to each node $i\in E_\alpha$ allows us to decompose the random walk into three steps:
\begin{enumerate}
 \item A random walker on node $i$ chooses an incident hyperedge $E_\alpha$ proportional to its weight,
 \begin{equation}
\label{eq:Q}
 Q_{\alpha i}=\frac{\omega(E_\alpha)e_{i\alpha}}{\sum_\beta \omega(E_\beta)e_{i\beta}}\, .
\end{equation}
 \item The random walker picks a node $j\neq i$ of hyperedge $E_\alpha$ proportional to its weight, 
 \begin{equation}
\label{eq:R}
 R_{j\alpha}=\frac{\gamma_{E_\alpha}(j)e_{j\alpha}}{\sum_{\ell\neq i}\gamma_{E_\alpha}(\ell)e_{\ell \alpha}}\, .
\end{equation}
 \item The random walker moves to node $j$ with probability 
 \begin{equation}
\label{eq:T}
T_{ij}=\sum_\alpha R_{j\alpha}Q_{\alpha i}\, .
\end{equation}
\end{enumerate}
\end{svgraybox}
Several random-walk processes can be defined with various node and hyperedge weights.
With hyperedge size as a proxy for the higher-order interactions' nature, we consider the transition probability 
\begin{equation}
\label{eq:Tij4}
T^{(\sigma)}_{ij}=\frac{K^{(\sigma)}_{ij}}{\sum_{{\ell\neq i}} K^{(\sigma)}_{i\ell}}\quad\forall i\neq j 
\quad \text{and} \quad
T^{(\sigma)}_{ii}=0\, ,
\end{equation}
with
\begin{equation}
\label{eq:khij}
K^{(\sigma)}_{ij}=\sum_{\alpha}(B_{\alpha \alpha}-1)^{\sigma}e_{i\alpha}e_{j\alpha}\quad\forall i\neq j
\quad \text{and} \quad
K^{(\sigma)}_{ii}=0\, ,
\end{equation}
for some real $\sigma$, biases random walks to hyperedges depending on their size \cite{carletti2020random,carletti2020randomwalks}.
By setting $\omega(E_\alpha)=(B_{\alpha \alpha}-1)^{\sigma+1}$ and $\gamma_{E_\alpha}(j)=(B_{\alpha \alpha}-1)$ for all $j\in E_{\alpha}$, Eq.~\eqref{eq:Tij4} takes the form of Eq.~\eqref{eq:T}. Indeed, from
\begin{equation*}
\sum_{\ell\neq i} K^{(\sigma)}_{i\ell} = \sum_{\ell\neq i}  \sum_{\beta}(B_{\beta \beta}-1)^{\sigma}e_{i\beta}e_{\ell\beta}=\sum_{\beta}(B_{\beta \beta}-1)^{\sigma+1}e_{i\beta}\, ,
\end{equation*}
where we used $\sum_{\ell\neq i}e_{\ell\beta}=B_{\beta \beta}-1$, we can conclude
\begin{align*}
T^{(\sigma)}_{ij}&=\frac{1}{\sum_{\beta}(B_{\beta \beta}-1)^{\sigma+1}e_{i\beta}}\sum_{\alpha}(B_{\alpha \alpha}-1)^{\sigma}e_{i\alpha}e_{j\alpha}\\
&=\frac{1}{\sum_{\beta}(B_{\beta \beta}-1)^{\sigma+1}e_{i\beta}}\sum_{\alpha}(B_{\alpha \alpha}-1)^{\sigma+1}e_{i\alpha}\frac{e_{j\alpha}}{B_{\alpha \alpha}-1}\\
&=\sum_\alpha Q_{\alpha i} R_{j\alpha}\, .
\end{align*}

For large positive values of $\sigma$, hyperedges with many nodes contribute more to $T_{ij}^{(\sigma)}$ and guide the random process. For large negative values of $\sigma$, the large hyperedges' contributions are negligible and only the small hyperedges drive the random walk process. In this way, $\sigma$ is a hyperedge-size bias parameter.

When $\sigma=0$, the random walk moves on the so-called clique-reduced multigraph, where each pair of nodes is connected by a number of edges equal to the number of hyperedges containing that pair in the hypergraph. The transition matrix takes the form
\begin{equation*}
 T^{(0)}_{ij}=\frac{K^{(0)}_{ij}}{\sum_{{\ell\neq i}} K^{(0)}_{i\ell}}=\frac{A_{ij}}{\sum_{{\ell\neq i}} A_{i\ell}} \, ,
\end{equation*}
where we used the hyperadjacency matrix $A_{ij}=\sum_{\alpha}e_{i\alpha}e_{j\alpha}$.
The clique reduced multigraph is different from the projected network obtained by associating to each hyperedge a clique of the same size. The latter can be interpreted as the unweighted version of the clique-reduced multigraph.

The matrix $K_{ij}^{(\sigma)}$ given by Eq.~\eqref{eq:khij} can be considered as the weighted adjacency matrix of an undirected network.  With the associated Laplace operator 
\begin{equation}
\label{eq:rwLap}
L^{(\sigma)}_{ij}=  \delta_{ij}-T^{(\sigma)}_{ij} = \delta_{ij}-\frac{K^{(\sigma)}_{ij}}{\sum_{\ell \neq i} K^{(\sigma)}_{i\ell}}\, ,
\end{equation}
the continuous-time random walk is
\begin{equation}
\dot{p}_i(t)=-\sum_j p_j(t)L^{(\sigma)}_{ji}\, ,
\label{eq:sigmaL}
\end{equation}
and the continuous-time transition matrix, which forwards the continuous-time random-walk process by time $t$, is
\begin{align}\label{eq:continuoustransition}
\bm{T}^{(\sigma)}(t) = e^{-t\bm{L}^{(\sigma)}}\, .
\end{align}

Based on this projection, we can invoke standard results about random walks and in particular prove that the stationary state~\footnote{To lighten the notation, we do not show explicitly the dependence on $\sigma$ on the asymptotic distribution $\bm{\pi}$.}
\begin{equation}
\pi_j =\frac{d_j^{(\sigma)}}{\sum_{\ell} d_{\ell}^{(\sigma)}}\,,
\label{eq:statnorm}
\end{equation}
where $d_j^{(\sigma)}=\sum_{\ell\neq j}K_{j\ell}^{(\sigma)}$ is the strength of node $j$ in the weighted graph. This quantity is an immediate generalisation of the standard node degree that accounts for different hyperedge sizes. 

The standard formulation of the map equation uses a discrete-time random-walk process. For discrete time steps $k$, the discrete-time version of Eq.~\eqref{eq:ctrw} takes the form 
\begin{align}\label{eq:dtrw}
    p_i(k+1) = \sum_j p_j(k)T_{ji}\,,
\end{align}
with the same stationary distribution as for the continuous-time process.

\section{Flow-based community detection in hypergraphs}

To illustrate how the flow-based community-detection methods known as Markov stability and the map equation identify communities in hypergraphs, we explain their disparate machinery and illustrate with a schematic hypergraph.

\subsection{Markov stability}
\label{ssec:MarkovStab}

The Markov stability is a quality function for partitioning a network into communities based on the persistence of random-walk flows inside a group of nodes.

Consider an ergodic random walk process in its stationary state. The Markov stability \cite{lambiotte2014random,Delvenne2010} of a partition at time $t$ is defined as the difference between the probability of a random walker to be in the same community at time $0$ and at time $t$ and the analogous quantity computed once the system settles in the stationary state.
At a given time $t$, Markov stability is large when random walkers are unlikely to have escaped their initial community. Because the process is assumed to be ergodic and knowledge of the initial conditions is lost asymptotically, the second probability is equal to the probability of two randomly chosen walkers residing in the same community.

\clearpage
\begin{svgraybox}
\textbf{Markov stability and autocovariance of the walker signal}\\
The Markov stability is equivalent to the autocovariance of a signal encoding the sequence of communities visited by a random walker in the stationary state.

Starting from the continuous-time random walk in Eq.~\eqref{eq:ctrw} or Eq.~\eqref{eq:sigmaL} and with a partition encoded by an $n \times C$ indicator matrix $\bm C$, we assign the values $X_\alpha$ ($\alpha=1,...,\mathcal{C}$) to the vertices of each community.
The autocovariance of the sequence of values $X(t)$ is
\begin{equation}
\mathrm{cov}\left[X(0)X(t)\right] = \langle X(0)X(t)\rangle- \langle X(0)\rangle \langle X(t)\rangle=\bm{X}^{\top} \bm{R}(t,\bm C) \bm{X}\, ,
\end{equation}
where $\langle X(t)\rangle$ is the expectation of the random variable $X(t)$, $\bm{X}$ is the $1\times \mathcal{C}$ column vector of labels assigned to the $\mathcal{C}$ communities and $\bm{R}(t,\bm C)$ is the $\mathcal C \times \mathcal C$ clustered covariance matrix,
\begin{align}\label{olo}
\bm{R}(t,\bm C)= \bm{C}^\top\left[\bm{\Pi} \exp(-t\bm{L})-\bm{\pi} \bm{\pi}^\top \right]\bm{C}\, ,
\end{align}
where $\bm{\Pi}= \text{diag}(\bm{\pi})$ is a diagonal matrix encoding the stationary distribution.
This $\mathcal{C}\times \mathcal{C}$ matrix does not depend on our arbitrary choice for the values $X_\alpha$. 

Since $(\bm{\Pi} \exp(-t\bm{L})_{ij}$ measures the flow of probability from node $i$ to node $j$ over time $t$, the first term of Eq.~\eqref{olo} measures the probability flow between two communities $\alpha$ and $\beta$ over a time $t$.  In a partition with a strong assortative community structure, the probability flows should be captured for long times within communities, and large entries in $\bm{R}(t,\bm C)$ should be concentrated on its diagonal. Following this reasoning, the Markov stability $r(t,\bm C)$ is defined as the trace of the clustered autocovariance matrix~\cite{Delvenne2010,Delvenne2013}: 
\begin{equation}
\label{eq:stability_disc}
r(t,\bm C) =  \Tr \left[\bm{R}(t,\bm C)\right]\, .
\end{equation}
\end{svgraybox}

The Markov stability quantifies the quality of a partition at different time scales.
For every $t$, maximising $r(t,\bm C)$ gives the best partition of the network into communities, resulting in a sequence of optimal partitions at different times. Time acts as a resolution parameter that enables us to tune the typical size of the communities in the optimal partition, as longer times typically lead to fewer and larger communities.
Markov stability shows connections with several concepts related to community detection. For small $t$, the first order of its Taylor expansion recovers, up to a constant, a parametric generalisation\cite{reichardt2006statistical} of Newman-Girvan modularity.

 \subsubsection{Markov stability for hypergraphs}
\label{ssec:MShyper}

The random walk process above defined in Eq.~\eqref{eq:sigmaL} corresponds to a random walk on a weighted undirected network. The link weights are self-consistently defined starting from the process on the hypergraph and thus taking into account the multibody interactions encoded by the size of the hyperedges.
Exploiting this relation, we can optimise  Markov Stability on the  weighted network to partition the hypergraph nodes into communities.
Consider again a partition of the nodes of a hypergraph into $\mathcal{C}$ non-overlapping communities, encoded by the $n \times \mathcal{C}$ indicator matrix $\mathbf{C}$. Expanding Eq.~\eqref{eq:stability_disc}, the corresponding Markov stability is explicitly given by
\begin{equation}
\label{eq:r}
r(t,\mathbf{C})=\Tr \left[ \mathbf{C}^{\top}\left(\mathbf{\Pi} e^{-t\mathbf{L}^{(\sigma)}} - \bm{\pi}^{\top}\bm{\pi}\right)\mathbf{C}\right] \, ,
\end{equation}
where $\bm{\pi}$ is the asymptotic solution of Eq.~\eqref{eq:statnorm}, $\mathbf{\Pi}$ is the diagonal matrix containing $\bm{\pi}$ on the diagonal, $\mathbf{L}^{(\sigma)}$ is the random-walk Laplace matrix and $\bm{\pi}^{\top}\bm{\pi}$ is the matrix whose $(i,j)$ entry is $\pi_i\pi_j$. 

As an illustration, we compute the Markov stability for a schematic hypergraph where nodes have two features represented by letters ($A$, $B$ or $C$) and numbers ($1$ or $2$) (Fig.~\ref{fig:hypergraphRepresentations}). The hypergraph has six nodes, $\{ A_1, A_2, B_1, B_2, C_1,C_2\}$, and five hyperedges, three of size two connecting nodes with the same letter, and two of size three connecting nodes with the same number.
With $\mathbf{K}^{(\sigma)}$ defined in Eq.~\eqref{eq:khij}, the hyperedges of size two have a weight of $1^\sigma$ to the associated link, whereas the hyperedges of size three have a weight of $2^\sigma$. The corresponding transition probabilities in Eq.~\eqref{eq:Tij4} are
\begin{align*}
\label{eq:transprob2feat}
 &T_{X_1,X_2}=\frac{1}{2\times2^\sigma+1}\, ,\\ &T_{X_1,Y_1}=\frac{2^{\sigma}}{2\times2^\sigma+1}\, ,\\
 &T_{X_1,Y_2}=0\, ,\\
 &T_{X_1,X_1}=0\, ,\\
 &T_{X_2,X_2}=0
\end{align*}
for all $X,Y$ in $\{A, B, C\}$.

When $\sigma=0$, the transition probabilities are $T_{X_1,X_2}=T_{X_1,Y_1}=1/3$. Thus, a walker remains twice as likely in the same $3$-hyperedge than in a $2$-hyperedge. The Markov stability thus returns two modules for sufficiently large  Markov times ($t>1$). For $\sigma=1$, $T_{X_1,X_2}=1/5$ and $T_{X_1,Y_1}=2/5$. Then, a walker is four times more likely to remain in the same $3$-hyperedge than in a $2$-hyperedge. For $\sigma=-1$, $T_{X_1,X_2}=1/2$ and $T_{X_1,Y_1}=1/4$, and the probability to stay in the $3$-hyperedge is the same as leaving it.
In the limit as $\sigma \to \infty$,
\begin{equation*}
 \lim_{\sigma\rightarrow \infty}T_{X_1,X_2}=0\, ,  \lim_{\sigma\rightarrow \infty}T_{X_1,Y_1}=1/2\quad \forall X,Y\in\{A,B,C\}\, ,
\end{equation*}
hops among nodes with the same number are strongly favoured and a walker remains for a long time in the same $3$-hyperedge. In the other limit,
\begin{equation*}
 \lim_{\sigma\rightarrow -\infty}T_{X_1,X_2}=1\, ,  \lim_{\sigma\rightarrow -\infty}T_{X_1,Y_1}=0\quad \forall X,Y\in\{A,B,C\}\, 
\end{equation*}
a walker spends longer time in the $2$-hyperedges. In the first case, Markov stability favours two communities for sufficiently large Markov times, while it favours three communities in the second case (Fig.~\ref{fig:phase}).

\subsection{The Map Equation}

Like Markov stability, the map equation is a quality function for partitioning a network into communities based on the persistence of random-walk flows inside groups of nodes. But instead of using autocovariance, the map equations casts the problem of finding flow-based communities in networks into a minimum-description-length problem~\cite{rosvall2008maps}.

\begin{svgraybox}
\textbf{The map equation} measures, in bits, the optimal codelength $L$ per step of a discrete random walk on a network for a given node partition $\mathsf{M}$ with $C$ modules. When all nodes are in the same module, the map equation is simply the Shannon entropy $H$ of the node-visit rates $\mathcal{P} = \{\pi_i\}$. 

In partitions with more than one module, the map equation combines within and between module codelengths for describing flows within and between modules.
For modules $\alpha=1,\ldots,C$ with 
\begin{align}
\begin{tabular}{@{}rl@{}}
entry flow rates & $q_{\alpha\curvearrowleft}=\sum_{i \notin \alpha, j \in \alpha}\pi_i T_{ij}$,\\
exit flow rates & $q_{\alpha\curvearrowright}=\sum_{i \in \alpha, j \notin \alpha}\pi_i T_{ij}$,\\
entry flow rate random variable & $\mathcal{Q} = \{q_{\alpha\curvearrowleft}\}$\\
with total flow rate & $q_{\curvearrowleft} = \sum_{\alpha}q_{\alpha\curvearrowleft}$,\\
exit and node-visit rate random variables & $\mathcal{P}_\alpha = \{q_{\alpha\curvearrowright}, \pi_{i \in \alpha} \}$ \\
with total flow rate & $p_{\alpha\circlearrowright} = q_{\alpha\curvearrowright} + \sum_{i \in \alpha}\pi_i$,\nonumber
\end{tabular}
\end{align}
the map equation takes its general two-level form
\begin{align}
L(\mathsf{M}) &= q_{\curvearrowleft}H(\mathcal{Q}) + \sum_\alpha p_{\alpha\circlearrowright} H(\mathcal{P}_\alpha). \label{eq:twolevelmapequation}
\end{align}
The first term is the codelength for between-module movements, followed by the sum of codelengths for within-module movements over all modules.
\end{svgraybox}

This standard formulation of the map equation encodes every transition of a random walker. For Markov times other than 1, the map equation uses a linearised continuous-time transition matrix,
\begin{align}\label{eq:linearizedtransition}
\tilde{\bm{T}}(t) =
\begin{cases}
    (1-t)\bm{I} + t\bm{T}     & t < 1\\
    t\bm{T}              & t \ge 1,
\end{cases}
\end{align}
which captures Markov times below 1 with self-links and Markov times above 1 with transition rates proportional to the average rate of the underlying Poisson process \cite{kheirkhahzadeh2016efficient}. At Markov time 1, it recovers the discrete-time process in Eq.~\eqref{eq:dtrw}.
Unlike the exponential of the Laplacian, the linearisation keeps the transition matrix sparse also for Markov times larger than zero.\footnote{Markov stability can also apply the linearisation to speed up calculations \cite{lambiotte2014random}, but here we use the exponential of the Laplacian for Markov stability in all examples.}

The linearisation is appealing for the map equation because the node-visit rates $\pi_i$ remain the same for all Markov times $t$ -- since the relative visit rates at steady state do not depend on how often the visits are sampled -- and the module exit and entry rates change linearly -- since the number of random walkers moving along any link between nodes during time $t$ is directly proportional to $t$. Therefore, we can define 
\begin{align}
q_{\alpha\curvearrowleft}(t) &= t q_{\alpha\curvearrowleft}\\
q_{\alpha\curvearrowright}(t) &= t q_{\alpha\curvearrowright}.
\end{align}
With time dependence on all variables in Eq.~\eqref{eq:twolevelmapequation} that depend on the module entry or exit flow rates, the map equation for Markov time $t$ takes the form
\begin{align}\label{eq:mapeqmarkovt}
L(\mathsf{M},t) &= q_{\curvearrowleft}(t)H(\mathcal{Q}(t)) + \sum_\alpha p_{\alpha\circlearrowright}(t) H(\mathcal{P}_\alpha(t)).
\end{align}

While the standard formulation of the map equation can be applied directly to the continuous-time transition matrix for various Markov times in Eq.~\eqref{eq:continuoustransition}, deriving the matrix is computationally expensive for large networks. For long Markov times, it also generates dense networks with longer clustering times. In contrast, the map equation for Markov time $t$ in Eq.~\eqref{eq:mapeqmarkovt} has no overhead compared with the standard map equation. From a flow modelling perspective, the map equation for Markov time $t$ counts each time a random walker moves across a modular boundary during time $t$. Instead, the standard map equation applied to the continuous-time transition matrix counts at most one boundary-crossing since it only considers the final position after time $t$. For short Markov times, the two approaches are similar. Bur for longer Markov times, the map equation for Markov time $t$ generally favours solutions with fewer modules \cite{kheirkhahzadeh2016efficient}.

To identify the optimal partition -- the one that compresses the modular description the most -- we use the community-detection algorithm Infomap \cite{edler2017mapping} available on \url{www.mapequation.org}. Infomap is to the map equation what the Louvain \cite{blondel2008fast} and the Leiden \cite{traag2019louvain} methods are to the objective function modularity \cite{newman2004finding}, which favours partitions with a high internal density of links compared with a statistical null model. Infomap uses a similar search algorithm as the Leiden method but tries to find the node assignment that minimises the map equation's codelength. 
Infomap also finds deeper hierarchical partitions -- from top-level supermodules with multiple levels of submodules down to leaf-level modules containing the nodes -- if such multilevel solutions give higher modular compression \cite{rosvall2011multilevel}. The search algorithm works on many network types, including weighted, directed, bipartite \cite{blocker2020mapping}, and multilayer \cite{de2015identifying} networks. Infomap can also identify overlapping multilevel communities in higher-order network flows modelled with so-called memory networks \cite{edler2017mapping}. Recent work introduces a Bayesian estimate of the map equation for identifying flows on sparse networks with missing links \cite{smiljanic2020mapping}, but here we focus on the standard map equation and identify hard two-level partitions in weighted unipartite networks obtained with the hypergraph projection described in section \ref{sec:rwh}. 

\subsubsection{The map equation for hypergraphs}

We have previously derived unipartite, bipartite, and multilayer network representations of hypergraph flows and analysed their different advantages when mapping flows on hypergraphs with Infomap \cite{eriksson2021mapping}. We found that when the research question does not require hyperedge assignments from using bipartite networks or overlapping modules from using multilayer networks, a unipartite network representation like the one we use here provides a good trade-off between speed and ability to reveal modular regularities.

We illustrate the map equation for hypergraphs with the schematic example in Fig.~\ref{fig:hypergraphRepresentations} for which the stationary distribution of the random walk is independent of $\sigma$ such that $\pi_{i}^{(\sigma)}=\pi_{i}$ for all nodes $i$. The one-module codelength is
\begin{align}
L(\mathsf{M}_1, t) &= H(\mathcal{P})\\
&= H(\pi_{A_1},\pi_{A_2},\pi_{B_1},\pi_{B_2},\pi_{C_1},\pi_{C_2}) \nonumber \\
& = 2.58 \text{ bits for all values of $\sigma$ and $t$.} \nonumber
\end{align}

When a network has modular regularities, a partition captures the modular flows when the random walker spends long times within the modules with few transitions between them.
The codelength is shorter than in the one-module solution because the information required to specify a random walker's position in a module decreases with its size.
But for partitions with too many modules, the information required for describing between-module movements exceeds the gain from using small modules. 
The optimal partition has the shortest codelength.
Its node assignment best captures the modular regularities of flows on the network.

Using the three-module solution in Fig.~\ref{fig:hypergraphRepresentations}(b), the codelengths for the  different hyperedge-size biased random walks parametrised with $\sigma$ are
\begin{align}\label{eq:toymapeqthreemodules}
L(\mathsf{M}_3, t) &=q_\curvearrowleft H(q_{1\curvearrowleft},q_{2\curvearrowleft},
q_{3\curvearrowleft})\\
&\mathrel{\hphantom{=}}+(q_{1\curvearrowright}+\pi_{A_1}+\pi_{A_2})
H(q_{1\curvearrowright},\pi_{A_1},\pi_{A_2})\nonumber\\
&\mathrel{\hphantom{=}} +(q_{2\curvearrowright}+\pi_{B_1}+\pi_{B_2}) H(q_{2\curvearrowright},\pi_{B_1},\pi_{B_2})\nonumber\\
&\mathrel{\hphantom{=}}+(q_{3\curvearrowright}+\pi_{C_1}+\pi_{C_2})
H(q_{3\curvearrowright},\pi_{C_1},\pi_{C_2})\nonumber\\
&\raisebox{9pt}{=}
\begin{tabular}{lcccl}
\ldelim\{{5}{4mm} & $\bm{1.00}$ & $\bm{1.00}$ & $\bm{1.00}$ & \text{ bits for $\sigma \to -\infty$,}\\
& $\bm{2.30}$ & 3.17 & 4.58 & \text{ bits for $\sigma = -1$,} \\
& 2.61 & 3.67 & 5.41 & \text{ bits for $\sigma = 0$,}\\
& 2.84 & 4.05 & 6.04 & \text{ bits for $\sigma = 1$, and}\\
& $\underbrace{\vphantom{8pt} 3.17}_{t=0.5}$
& $\underbrace{\vphantom{8pt} 4.58}_{t=1}$
& $\underbrace{\vphantom{8pt} 6.92}_{t=2}$
& \text{ bits for $\sigma \to \infty$.} 
\end{tabular} \nonumber
\end{align}
This three-module letter-based solution gives shorter codelengths than the one-module solution for small $\sigma$-values and short Markov times, indicated by bold numbers in Eq.~\eqref{eq:toymapeqthreemodules}. For these values, the random walker spends long times within the three letter-based modules with few transitions between.

Using the two-module solution in Fig.~\ref{fig:hypergraphRepresentations}(d), the codelengths for the  different hyperedge-size biased random walks parametrised with $\sigma$ are
\begin{align}\label{eq:toymapeqtwomodules}
L(\mathsf{M}_2, t) &=q_\curvearrowleft H(q_{1\curvearrowleft},q_{2\curvearrowleft})\\
&\mathrel{\hphantom{=}}+(q_{1\curvearrowright}+\pi_{A_1}+\pi_{B_1}+\pi_{C_1})
H(q_{1\curvearrowright},\pi_{A_1},\pi_{B_1},\pi_{C_1})\nonumber\\
&\mathrel{\hphantom{=}} +(q_{2\curvearrowright}+\pi_{A_2}+\pi_{B_2}+\pi_{C_2})
H(q_{2\curvearrowright},\pi_{A_2},\pi_{B_2},\pi_{C_2})\nonumber\\
&\raisebox{9pt}{=}
\begin{tabular}{lcccl}
\ldelim\{{5}{4mm} & 3.46 & 4.58 & 6.34 & \text{ bits for $\sigma \to -\infty$,}\\
& 2.74 & 3.46 & 4.58 & \text{ bits for $\sigma = -1$,} \\
& $\bm{2.44}$ & 3.00 & 3.87 & \text{ bits for $\sigma = 0$,}\\
& $\bm{2.17}$ & $\bm{2.56}$ & 3.19 & \text{ bits for $\sigma = 1$, and}\\
& $\underbrace{\vphantom{8pt} \bm{1.58}}_{t=0.5}$
& $\underbrace{\vphantom{8pt} \bm{1.58}}_{t=1}$
& $\underbrace{\vphantom{8pt} \bm{1.58}}_{t=2}$
& \text{ bits for $\sigma \to \infty$.} 
\end{tabular} \nonumber
\end{align}
This two-module number-based solution gives shorter codelengths than the one-module solution for large $\sigma$-values and short Markov times, again indicated by bold numbers in Eq.~\eqref{eq:toymapeqtwomodules}. For these values, the random walker instead spends long times within the two number-based modules so the codelength savings from using smaller modules exceed the extra cost from switching modules.

\subsection{Comparing Markov stability with the map equation}

\begin{figure}[htb]
    \centering
    \includegraphics[width=0.6\textwidth]{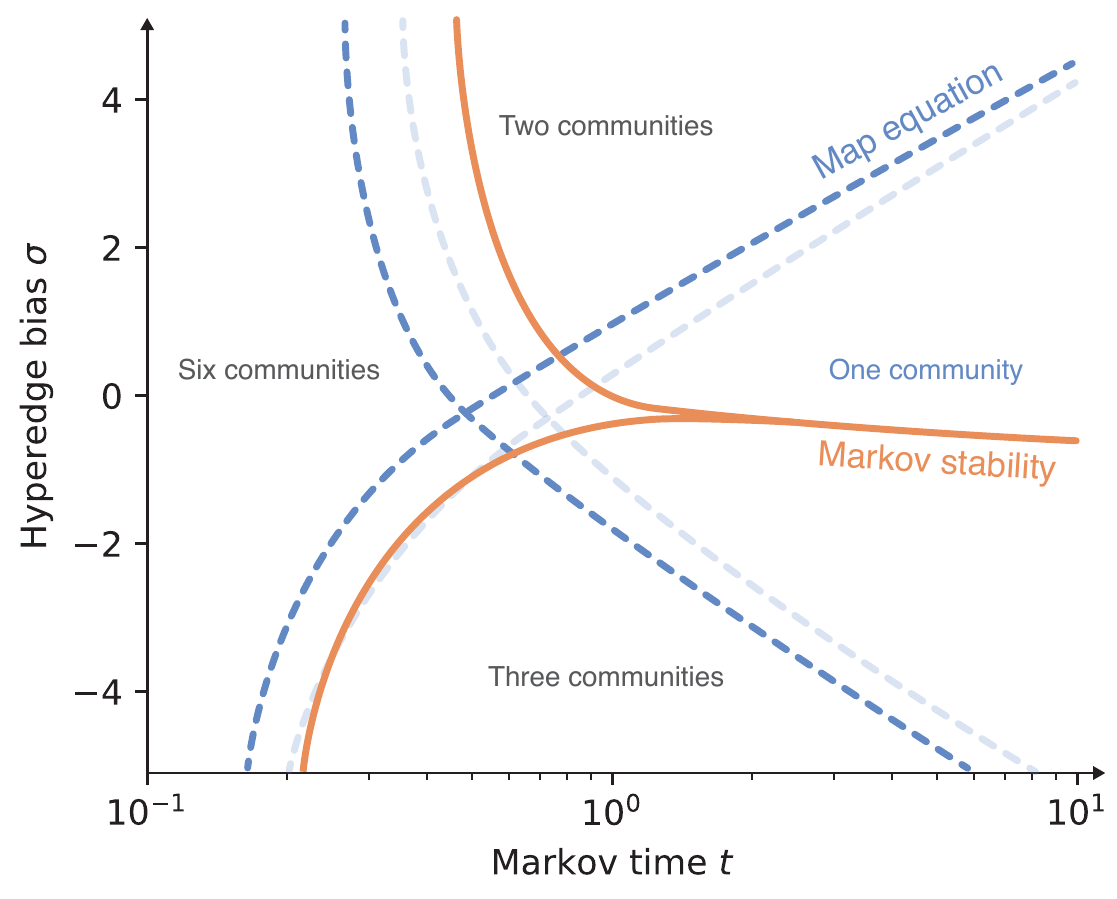}
    \caption{Optimal solutions of the schematic hypergraph in Fig.~\ref{fig:hypergraphRepresentations} for different hyperedge biases and Markov times. Solutions from Fig.~\ref{fig:hypergraphRepresentations}(b-d) and the singleton solution with six communities. Light dashed line for the map equation applied to a continuous-time process.}
    \label{fig:phase}
\end{figure}

While both Markov stability and the map equation identify flow-based communities in hypergraphs, they highlight different structures. Both methods prefer the solution with one node in each community for minimal Markov times when nearly all flow remains in the nodes. But the Markov time at which they highlight non-trivial solutions differs. For Markov stability this happens when more flows stay in larger communities than in singleton communities compared with respective stationary expectation. For the map equation, it happens when describing transitions between communities costs more than gained from describing within-community movements in singleton communities. Which one comes first depends on the network structure. For the schematic network in Fig.~\ref{fig:hypergraphRepresentations}, the map equation first prefers non-trivial solutions (Fig.~\ref{fig:phase}). But in the larger networks analysed in section \ref{sec:experiments}, Markov stability highlights non-trivial solutions for shorter Markov times than the map equation.

The map equation strictly prefers the one-community solution over a non-trivial solution for long Markov times. It prefers a two-community solution only when describing transitions between the communities costs no more than gained from describing within-community movements in smaller communities. The map equation's modular code structure with costs in bits for both community exits and entries has a regularising effect as long as the network is connected. In contrast, Markov stability prefers any solution that traps the slightest more flows within the communities for Markov time $t$ than expected after infinite time.
For example, only the map equation prefers the one-community solution for the longer Markov times included in Fig.~\ref{fig:phase}. As a result, Markov stability typically has a longer range of non-trivial solution and responds slower to changes in Markov time.

For the schematic hypergraph, both methods prefer two communities with flows biased to larger hyperedges and three communities when biased to smaller hyperedges (Fig.~\ref{fig:phase}). The map equation's continuous-time approximation slightly shifts the optimal solutions to longer Markov times in this case.  

\section{Experiments}\label{sec:experiments}

To illustrate how Markov stability and the map equation differ applied to real hypergraphs, we analysed a zoo hypergraph, a collaboration hypergraph, and a hypergraph from fossil records.

\subsection{A zoo hypergraph}

The zoo hypergraph has been built using data from the UCI Machine Learning Depository~\cite{Dua:2019}; it consists of $101$ animals nodes and $20$ features hyperedges such as fur, having a tail, and the number of legs. We identified communities for random-walk bias $\sigma =$~$-4$, $-2$, and $2$ as a function of Markov time for Markov stability using public code \cite{markovstability} and the map equation using Infomap \cite{infomap}. 

\begin{figure}[htb]
    \centering
    \includegraphics[width=\textwidth]{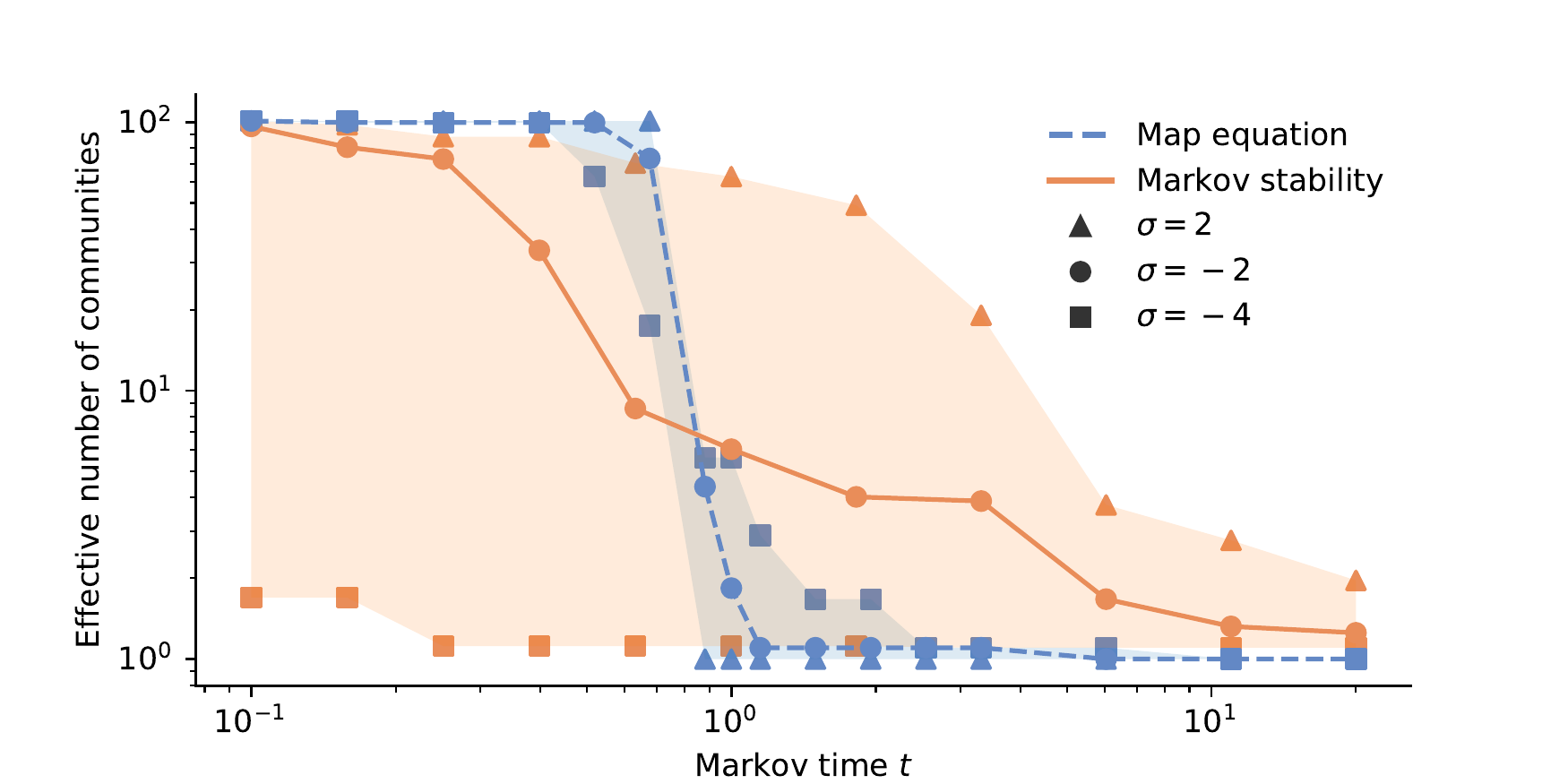}
    \caption{Effective number of communities in the zoo hypergraph for different hyperedge-size bias $\sigma$ and Markov time $t$. Filled areas between the highest and lowest number of effective communities for Markov stability (orange) and the map equation (blue).}
    \label{fig:zoo}
\end{figure}

The effective number of communities, measured as the perplexity of the relative community sizes $\mathcal{S}$, $2^{H(\mathcal{S})}$, decreases gradually for Markov stability and more abruptly for the map equation (Fig.~\ref{fig:zoo}). Already at Markov time $0.1$, Markov stability applied to the zoo hypergraph with hyperedge-size bias $\sigma = -4$ identifies less than two effective communities. The bias toward small hyperedges fragments the network into weakly connected components. The penalty term $-\bm{\pi}^{\top}\bm{\pi}$ in Eq.~\eqref{eq:r} with global flows makes Markov stability sense this large-scale network structure even if most flows -- more than 90 per cent in this case -- remain at the node for the short Markov time.

In contrast, the map equation prefers the singleton solution for the three tested hyperedge-size biases for Markov time up to $0.5$.  A short Markov time makes encoding each node in its singleton community cheap since most flows remain there in each step. The extra cost from describing movements in larger communities is higher than the gain from describing fewer transitions between communities until the Markov time approaches $1$. 

Markov stability identifies the largest communities with flows biased to small hyperedges for all Markov times. The map equation applied to a continuous-time process shows similar behaviour. But for the approximation shown in Fig.~\ref{fig:zoo}, it favours the one-community solution with flows biased to large hyperedges for Markov time about $0.9$ or higher. Encoding all nodes in the same community best compresses the interconnected flows biased to large hyperedges.

\begin{figure}[htb]
    \centering
    \includegraphics[width=\textwidth]{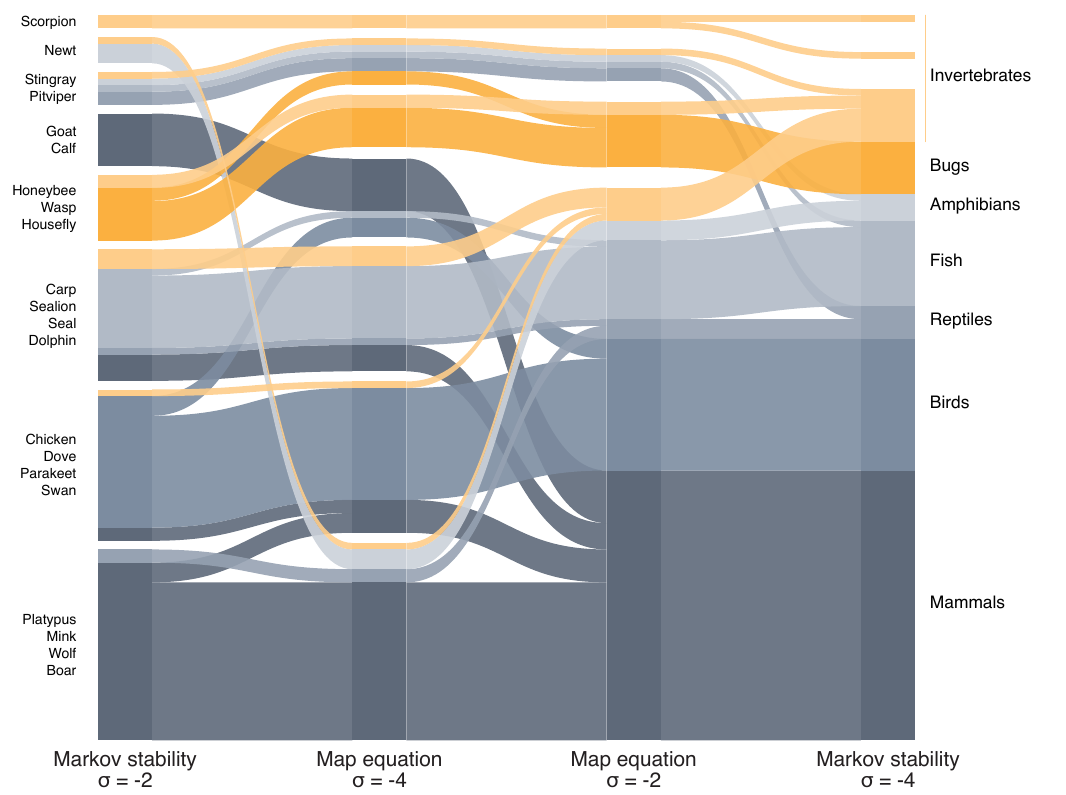}
    \caption{Alluvial diagram of the zoo hypergraph solutions for Markov time $t=1$ and different hyperedge biases $\sigma$. Each block represents a community with block height proportional to the number of nodes in the community. Stream fields connect blocks with shared nodes. Colours indicate animal classes.}
    \label{fig:zooAlluvial}
\end{figure}

The various community structures at Markov time $1$ illustrates that Markov stability and the map equation work differently. They identify different communities for similar flows and similar communities for different flows. Between $\sigma = -2$ and $-4$, Markov stability and the map equation find closest solutions for differing $\sigma$. Markov stability for $\sigma = -2$ and the map equation for $\sigma = -4$ find communities that best capture animal classes (Fig.~\ref{fig:zooAlluvial}). 

\subsection{A collaboration hypergraph}

We used a collaboration hypergraph compiled from the 734 references in the review paper \emph{Networks beyond pairwise interactions: Structure and dynamics} \cite{battiston2020networks,eriksson2021mapping}.
Hyperedges represent referenced articles and nodes represent their authors. Authors with multiple articles form connections between the hyperedges. We analysed the largest connected component with $361$ author nodes in $220$ hyperedges.

\begin{figure}[htb]
    \centering
    \includegraphics[width=\textwidth]{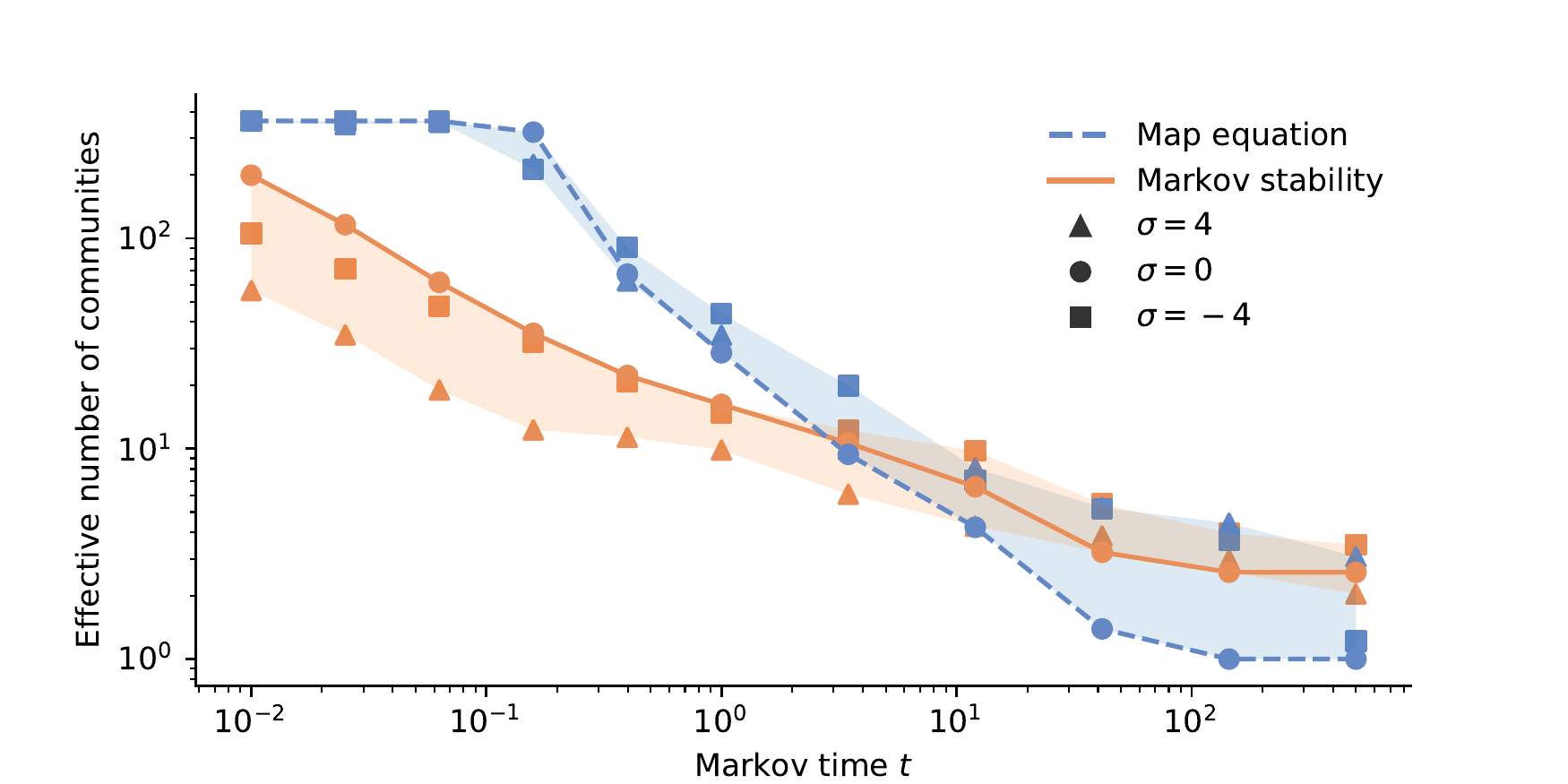}
    \caption{Effective number of communities in the collaboration hypergraph for different hyperedge-size bias $\sigma$ and Markov time $t$. Filled areas between the highest and lowest number of effective communities for Markov stability (orange) and the map equation (blue).}
    \label{fig:references}
\end{figure}

Again, changes in Markov time have a more gradual effect on Markov stability than on the map equation. But for the collaboration network, the map equation transitions from the singleton to the all-in-one solution over three orders of magnitude and identifies non-trivial solutions for Markov times over $100$ (Fig.~\ref{fig:references}). Different Markov times highlight different hierarchical levels in the hypergraph's hierarchical community structure with smaller communities nested in larger communities. The largest communities remain clear despite long walks that multiply flows tenfold between the communities.

\subsection{A fossil-record hypergraph}

We analysed a hypergraph representation of marine fossil animals from Cambrian (541 MY) to Cretaceous (66 MY) \cite{rojas_multiscale_2021}. Geological stages in the underlying sample-based occurrence data form hyperedges connecting all genera occurring at each stage. Genera occurring in multiple geological stages connect hyperedges. We weighted the hyperedges by dividing the number of samples where a genus occurs in a given geological stage by the total number of samples recorded at the stage. The assembled hypergraph comprises 77 geological stage hyperedges and 13,276 fossil genera nodes \cite{eriksson2021mapping}.

\begin{figure}[htb]
    \centering
    \includegraphics[width=\textwidth]{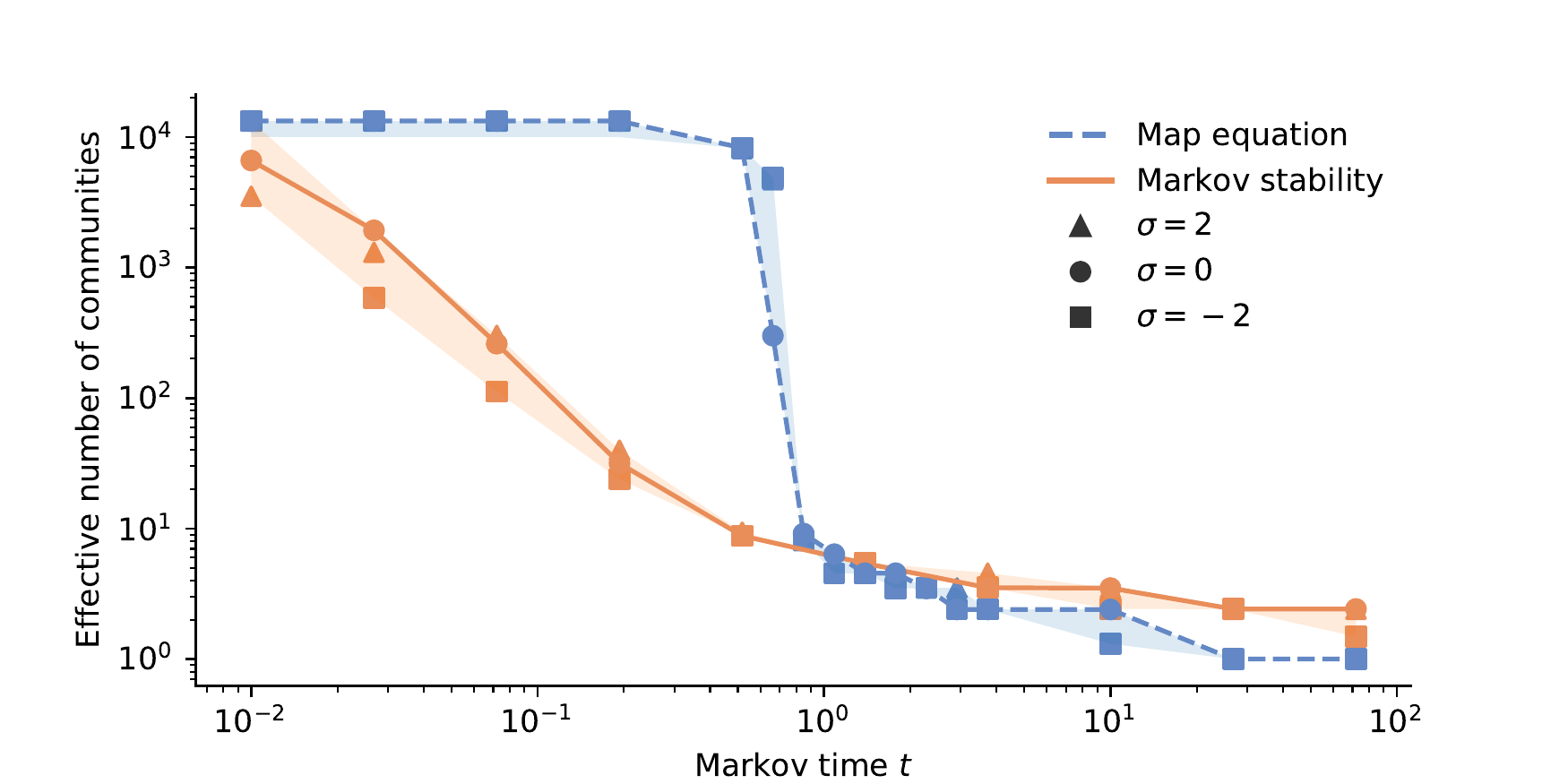}
    \caption{Effective number of communities in the fossil hypergraph for different hyperedge-size bias $\sigma$ and Markov time $t$. Filled areas between the highest and lowest number of effective communities for Markov stability (orange) and the map equation (blue).}
    \label{fig:fossilrecord}
\end{figure}

Once more, the effective number of identified communities decreases gradually for the Markov stability without reaching the all-in-one solution even at high Markov times (Fig.~\ref{fig:fossilrecord}). In contrast, the map equation solutions quickly transition from singleton to all-in-one communities around Markov time $1$. In the narrow range of Markov times with non-trivial solutions, the effective number of communities is two to five. For this substantially larger hypergraph, the extensive hyperedges form a weak community structure that dissolves for long Markov times and prevents modular compression.

\begin{figure}[h!]
    \centering
    \includegraphics[width=\textwidth]{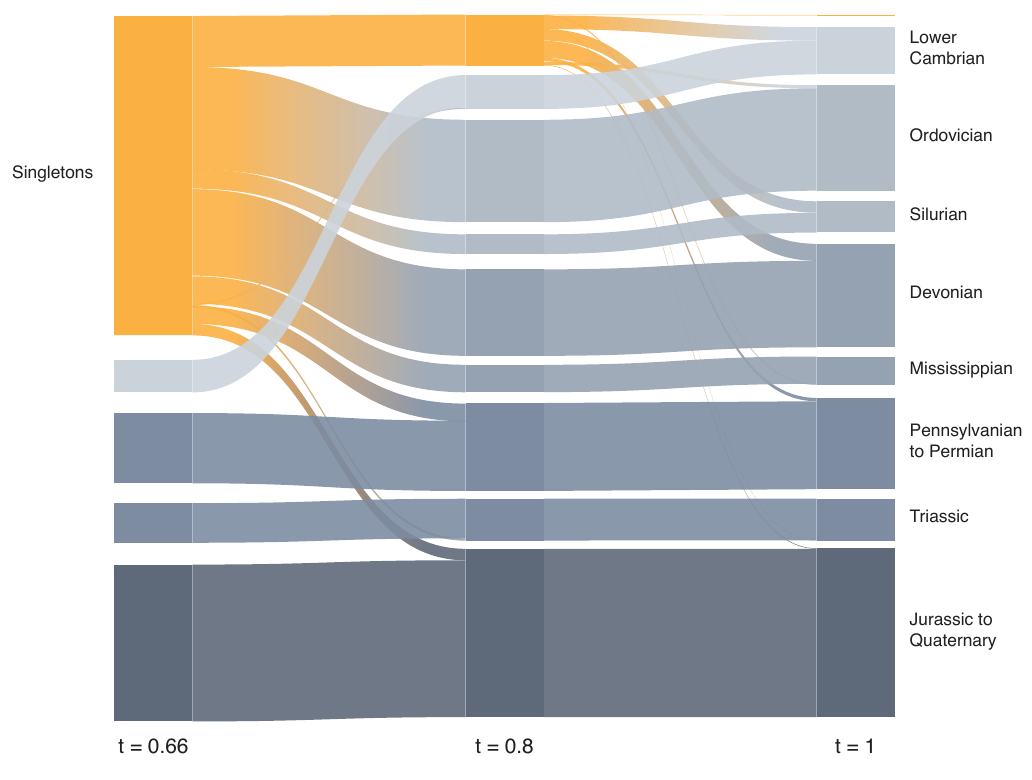}
    \caption{Alluvial diagram of the map equation's fossil hypergraph solutions for hyperedge-size bias $\sigma = 0$. Each block represents a community with block height proportional to the number of nodes in the community. Stream fields connect blocks with shared nodes. Singleton communities aggregated in orange blocks for clarity.}
    \label{fig:fossilAlluvialME}
\end{figure}

Non-trivial Infomap solutions reproduce the underlying temporal structure of the paleontological data with faunas organised into units of geological time (Fig.~\ref{fig:fossilAlluvialME}). Although fossil genera can occur in more than one of these large-scale temporal units, Infomap identifies successive global faunas that replace each other at their boundaries. Overall, faunas from geological periods are clustered together or combined into coarser temporal units at Markov time $1$. These salient structures appear from singletons without intermediate smaller structures for shorter Markov times.

\begin{figure}[H]
    \centering
    \includegraphics[width=\textwidth]{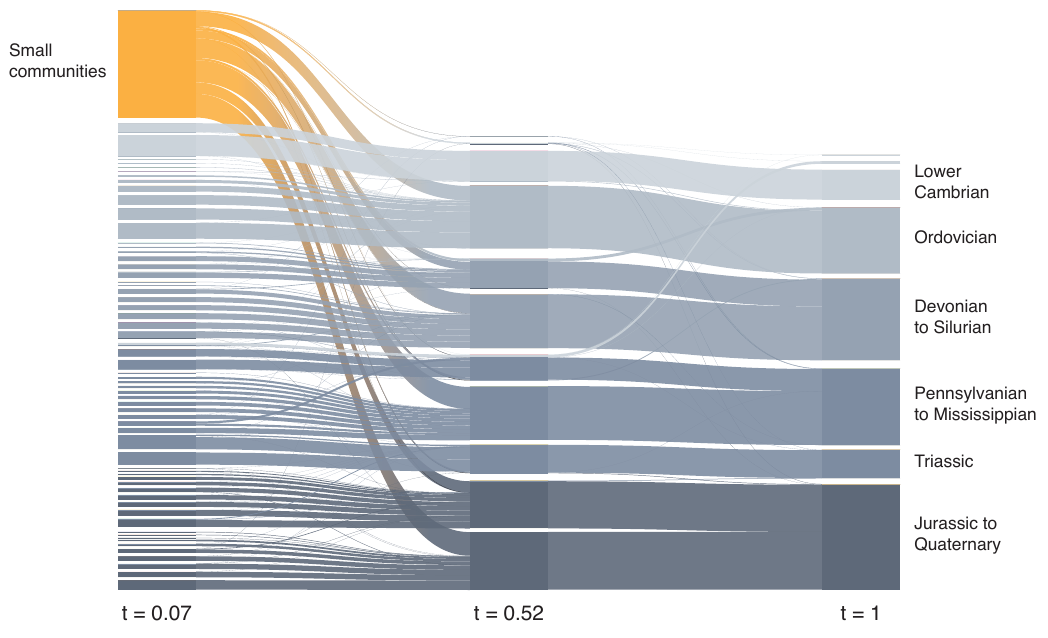}
    \caption{Alluvial diagram of Markov stability's fossil hypergraph solutions for hyperedge-size bias $\sigma = 0$. Each block represents a community with block height proportional to the number of nodes in the community. Stream fields connect blocks with shared nodes. Markov times chosen to match the effective number of communities in Fig.~\ref{fig:fossilAlluvialME} and small communities with fewer than five nodes aggregated in an orange block for clarity.}
    \label{fig:fossilAlluvialMS}
\end{figure}

Markov stability also delineates temporal faunas limited by geological units at Markov time around $1$. However, the temporal structure of the data cannot explain the numerous communities Markov stability identifies for short Markov times (Fig.~\ref{fig:fossilAlluvialMS}).

\section{Conclusions}
We have derived Markov stability and the map equation for a random-walk process on hypergraphs with hyperedge-size bias. Both methods identify communities where flows stay for long times. Still, with disparate machinery -- Markov stability measures overrepresentation of random walkers in communities where they started, whereas the map equation measures the modular description length in bits -- they capitalise on distinct flow-based structures. By comparing with the stationary expectation, Markov stability is more sensitive to hyperedge-size biases and gradually finds larger communities for longer Markov times. When increasing the Markov time, the map equation instead transitions more abruptly from identifying many small to few large communities. Compared with the influence from their disparate machineries, the map equation's approximation of a continuous-time process has a negligible effect.
Whereas the map equation identifies salient structures, Markov stability can identify communities of any size irrespective of the large-scale hypergraph structure. The question and hypergraph at hand decide which method identifies the flow-based communities that best elucidate the hypergraph's studied function.

\begin{acknowledgement}
A.E was supported by the Swedish Foundation for Strategic Research, Grant No.\ SB16-0089. A.R. and M.R.\ were supported by the Swedish Research Council, Grant No.\ 2016-00796
\end{acknowledgement}
%
%
%

\bibliographystyle{spmpsci}
\bibliography{biblio}
\end{document}